# Developmental Status and Perspectives for Tissue Engineering in Urology.

**Elcin Huseyn**
Azerbaijan State Oil and Industry
University, Baku, Azerbaijan
elcin.huseyn@asoiu.edu.az

**Abstract:** Tissue engineering technology and tissue cell-based stem cell research have made great strides in treating tissue and organ damage, correcting tissue and organ dysfunction, and reducing surgical complications. In the past, traditional methods have used biological substitutes for tissue repair materials, while tissue engineering technology has focused on merging sperm cells with biological materials to form biological tissues with the same structure and function as their own tissues. The advantage is that tissue engineering technology can overcome donors. Material procurement restrictions can effectively reduce complications. The aim of studying tissue engineering technology is to find sperm cells and suitable biological materials to replace the original biological functions of tissues and to establish a suitable in vivo microenvironment. This article mainly describes the current developments of tissue engineering in various fields of urology and discusses the future trends of tissue engineering technology in the treatment of complex diseases of the urinary system. The results of the research in this article indicate that while the current clinical studies are relatively few, the good results from existing animal model studies indicate good prospects of tissue engineering technology for the treatment of various urinary tract diseases in the future.

## 1. Introduction

Urinary system tissue damage and organ loss caused by degenerative diseases, trauma or tumors have always been problems that urology needs to solve. In the past, autologous tissues were used to replace them, such as genital skin, bladder mucosa, oral mucosa, etc., but often It can cause various complications such as hair growth, stenosis, stone formation, and diverticulum formation. Moreover, because the materials are taken from their own tissues, they are often limited by the source of the donor material, and complications at the donor site are prone to occur. In the case of end-stage organ failure, transplantation is usually used. Although transplantation technology is relatively mature, the results are not always satisfactory in terms of immune control and suppression of complications. In addition, the number of organ donors in the country and the world is far below the demand for transplantation, and the population is gradually aging, and the number of people with reduced tissue and organ function is increasing. If the length of the ureter and urethra is large, repair, operations are sometimes difficult to perform. Therefore, to solve these problems, it is necessary to find a way to replace or regenerate damaged organs and tissues. The development of regenerative medicine and tissue engineering allows people to further understand the mechanism of tissue regeneration and find ways to regenerate damaged tissues or organs. It is now possible to improve tissue regeneration capacity or use biomaterial scaffolds combined with stem cells to repair damaged tissues and organs, to restore the functions of tissues and organs and improve the quality of life of patients. This article reviews the achievements in basic research and clinical application of tissue engineering technology in urology in recent years, to provide direction and guidance for future research and innovation in this field.

## 2. Overview of urinary system tissue engineering

Tissue engineering is the science of constructing biological substitutes to maintain or repair damaged tissues and organs based on the principles of cell biology, materials science, and bioengineering.

The cell sources of urinary system tissue engineering mainly include:

- Original cells: obtained from oneself, cultured in vitro, and then used to repair damaged tissue.
- Urinary tract epithelial cells: Obtained through bladder biopsy and cell culture. The collection method is invasive, but it is currently available. Urinary epithelial cells are separated by urine or bladder washing.
- Autologous epithelial cells, oral keratinocytes, smooth muscle cells (used to restore the contraction and urination function of the bladder).
- Stem cells: bone marrow stem cells, adipose stem cells, urine-derived stem cells.

The scaffold materials for tissue engineering mainly include:

1. Synthetic polymers, such as biodegradable polymers: polyglycolic acid and polylactic acid-polyglycolic acid, both of which are biomaterial's composed of covalently bonded macromolecules. The main advantage of synthetic polymers is that they can produce organ structures of any shape in three-dimensional space, can be quantified, have good reproducibility, and are relatively low cost. Because it is an artificial material, there are no problems related to graft or autologous tissue collection damage. In addition, many characteristics of such synthetic polymers can be controlled, such as porosity and mechanical properties; and can be degraded through hydrolytic pathways, and the residual debris can be eliminated through metabolic pathways; it does not contain any molecular signals related to guiding cell activity and differentiation, so it is more conducive to tissue and organ repair.

2. Biologically derived scaffolds are decolourized tissues based on chemical and mechanical means, such as submucosa of the small intestine (SIS) and bladder acellular matrix (BAM). Due to the residual growth factors and extracellular matrix (ECM) proteins, they have the advantages like the inherent biological activity and mechanical properties of natural ECM. However, one of the main disadvantages of these stents is that the remaining protein composition and structure may be different from the body itself to be implanted. Most natural scaffolds come from pigs, so they may be a source of disease transmission. In addition, ethical issues will also affect its clinical use.

3. Self-assembled engineering, tissue. This method utilizes or combines the self-assembly characteristics of cells to construct a three-dimensional biological tissue structure. After decellularization and sterilization of biologically derived materials, the exogenous extracellular matrix material may still contain a considerable amount of DNA that may affect biocompatibility. The self-assembly method can produce tissues constructed by the cells themselves, in which the dense extracellular matrix is produced entirely by their own fibroblasts. The biggest advantage of this material is that it eliminates the influence of the biocompatibility of exogenous materials. By reducing the immune response, inflammation and fibrosis can be reduced, thereby increasing the success rate of surgery. In recent years, some of the techniques in these methods have been used in the reconstruction of urinary system tissues [1]. These techniques require cells to receive corresponding signals for proper differentiation, so that the engineered tissues for transplantation and the target tissues for replacement can be exhausted. It may be similar. Studying and simulating related signals is a difficult problem that needs to be overcome to apply these technologies to other tissues and organs.

In addition, nowadays 3D printing technology has been widely used in various fields, including nanoelectronics, medicine and tissue engineering. It can accurately combine different materials with different substrates, making it the best choice for carrying drugs and conducting personalized medical research. [2] Therefore, 3D printing technology has gradually become an important part of the construction of tissue engineering scaffolds.

### 3. Application of tissue engineering technology in kidney

The main components used to construct a functional kidney structure in tissue engineering technology are living cells, a scaffold system based on biometrics, biologically active factors, and appropriate microenvironments that promote cell behavior. On this basis, the natural healing ability of body regeneration is used to guide the growth of new organizations. The method is mainly to separate the donor tissue into individual cells. These cells are either directly implanted in the host body, or cultured and expanded and then implanted, or attached to a scaffold and expanded and then implanted [3].

As a good carrier for cell growth, collagen hydrogel has been widely used in the preparation of kidney tissue engineering scaffolds. The three-dimensional collagen scaffold was used to reconstruct three-dimensional kidney tissue in vitro using mixed newborn rat kidney cells, and it was found that the seed cells in the three-dimensional hydrogel scaffold could self-assemble into engineered kidney tissue containing renal tubules and glomerulus-like structures and be cultured in three dimensions. The cells can retain their phenotype, migration ability and albumin uptake function. Hydrogel hyaluronic acid (HA) can be used as a scaffold material for kidney tissue engineering, but actual research and application have shown that hydrogel scaffolds and other hydrogel polymer scaffolds are difficult to maintain due to low mechanical strength and physical shape. This results in a higher failure rate during use. To improve the mechanical strength, there are also studies combining extracellular matrix derivatives with synthetic biomaterials. This composite scaffold is considered a solid carrier in tissue engineering [4]. In terms of decellularized scaffolds, the researchers decellularized rat kidneys, implanted epithelial cells and endothelial cells to create cell-containing scaffolds, then perused the cell-containing scaffolds in a director, and finally implanted them in the body. The bioengineered kidney after transplantation can produce urine and remove metabolites in the body [5].

In addition to the above research, another main reason that bioengineered organs can achieve long-term good results in the body is a potency of blood vessels. In the case that the vascular matrix is not completely fused with endothelial cells, there is likely to be obvious thrombosis in the vascular system of the stent, which makes the revascularized structure lose its function. To solve this problem, Ko et al. [6] Adopted an endothelial cell inoculation method that can effectively cover the blood vessel wall of the decellularized pig kidney stent. The study of Lertkiatmongkol et al. [7] Showed that the combination of CD31 antibody and vascular matrix enables vascular endothelial cells to attach to blood vessels. CD31 is also known as platelet endothelial cell adhesion molecule-1 (PECAM-1/CD31).). Therefore, in future research, we can try to use endothelial cell inoculation method and combine with antibody to effectively improve the adhesion and retention of endothelial cells, and further investigate whether this method can make the engineered renal blood vessels unobstructed.

Machiguchi et al. [8] Used cell-cell interactions in conditioned media to generate neurons in the body for kidney repair. Their research proved the inhibitory effect of collecting duct cell matrix of vascular endothelial cells and renal tubular epithelial cells, as well as the stimulating interaction between vascular endothelial cells and renal tubular epithelial cells. Since the collecting duct cell matrix is like the renal tubular epithelial cell matrix, it can cause mesenchymal stem cells to differentiate into renal tubular epithelial cells. Therefore, renal tubular epithelial cells differentiated from mesenchymal stem cells induced by the collecting duct cell matrix can be injected into the rat renal cortex. Compared with unprotected cells, pre-implantation of cells with a small amount of gel complex for three-dimensional culture pretreatment can trigger the formation of more neoprene-like structures, indicating that the pre-treated kidney is injected into the renal cortex. Tubular epithelial cells may help repair dysfunctional kidney tissue. Zhang et al. [9] Used adipose stem cells to differentiate into mesoderm and implanted into stents, and then recellularized the renal artery and ureter. They found that the induced mesoderm cells differentiated into tubular cells and products more efficiently than adipose stem cells. This will provide a new idea for the differentiation and induction of seed cells in the future.

In order to make the above technology feasible in clinical transformation, other key technical problems need to be solved, including manufacturing a cell-free renal scaffold with clinical scale [10]; using clinically sourced cells to effectively recellularize the scaffold for reconstruction Fully functional kidney structure; reduce the cycle of cultivating engineered kidney; ensure that there is no serious thrombosis in the long-term implant. In addition, the clinical application of fully functional kidneys in kidney-specific diseases are still a challenging task. To achieve this goal, multidisciplinary research in this field should be carried out in a balanced manner, including the establishment of kidney disease models, preclinical research on the construction and application of tissue engineering kidneys, treatment methods for immune problems, and the regulation of innervation of implants. The function of the kidney is complex, and the structure is diverse. It is still difficult to completely reconstruct an engineered kidney that can be used for transplantation by tissue engineering in vitro. Therefore, it is necessary to continue to select and optimize to produce engineered kidneys in future research. The cells, scaffold materials and culture, the environment are used to make engineered kidneys that function closer to normal kidneys and can be used for transplantation.

### 4. Application of tissue engineering technology in ureter

Acellular matrix is a kind of tissue engineering ureteral scaffold, especially when it is planted with cells, it can increase the formation of blood vessels. Koch et al. [11] Proposed to use porcine acellular matrix cross-linked ureter as a scaffold for human arterial wall regeneration. The results of the study found that the smooth muscle cells inoculated within 2 weeks evenly filled the stent, making the stent a substitute for the ureter for transplantation. To improve the regeneration effect, Zhao et al. [12] Proposed the method of implanting mesenchymal stem cells between vascular endothelial cells to bridge the long arterial gap. The differentiation of mesenchymal stem cells can obtain multi-layered arterial tissue. In smooth muscle tissue stained with α-smooth muscle actin and smooth muscle myosin heavy chain, urothelial cytokeratin 20 and write-in 11 were positively expressed, which proves that this technology is useful in tissue engineering arterial repair. Application prospects clinically, the treatment methods for arterial stenosis or injury are mainly insertion of a double J tube, balloon expansion, repair, and skin fistula. The application of tissue engineering technology will greatly improve the quality of life of patients who have to fistula, such as the ureter that cannot be repaired or has long defects. Clinically, the standard method of replacement therapy is tubular small intestine replacement therapy or tongue mucosal replacement therapy, but due to many complications, it is difficult to achieve the expected clinical treatment effect. Another pioneering research is the use of veins instead of ureters [13]. The vein is a pipeline for the transportation of fluid in the human body. It has good elasticity and is very close to the ureter in function. However, the number of animals in this study is small, and the incidence of hydronephrosis is still increasing 3 months after transplantation, and the effect is not very satisfactory. Further studies need to have good long-term results before they have the chance to be applied in clinical practice in the future.

In the future application of tissue engineering technology, the concept of vein replacement in the above method can be combined, and the decellularized vein can be used as a stent to plant arterial epithelial cells and smooth muscle cells on the stent to achieve the function of replacing the ureter. The location of the ureter is deep, and the injuries are mostly iatrogenic. The patients with long stenosis are often older and have a higher tolerance to conventional treatment. Therefore, the applicability of tissue engineering technology for the treatment of arterial stenosis lacks effective clinical practice. Research.

## 5. Application of tissue engineering technology in the bladder

The materials for bladder tissue engineering mainly include:

1. Biomaterial's: natural biological materials made of collagen and alginate and cellular tissue matrices made of different types of tissues from pigs, such as the submucosa of the bladder, the submucosa of the small intestine, the dermis, Bladder, gallbladder, and amniotic fluid tissue, etc. [14].

2. Artificial materials: synthetic stents (polyethylene sponge, Teflon, etc.); synthetic polymers (such as poly-α-ester); silk-based materials. These materials have the advantages of non-toxicity, biodegradability, and easy adjustment of their structure. Among them, the silk-based sourcing has been removed from the silk biological material, and the human body's tolerance to the silk-based material is equivalent to that of the biological material, so the silk-based material can be used as a scaffold in tissue engineering technology. In addition, Shakhssalim et al. [15] Successfully applied electrospinning technology to prepare a polycaprolactone/polylactic acid scaffold for canine bladder wall replacement. Cell matrix was added to support seed detrusor smooth muscle cells, which can stimulate local in vivo growth of primary cells.

In terms of cell selection, in theory, autologous cells are the best choice for inducing tissue fusion and regeneration, which can effectively avoid serious immune responses. However, if urine-derived stem cells are collected at the source of autologous cells, they are not suitable for urinary diversion. The main reason for patients undergoing urinary diversion surgery is that most of the patients undergoing urinary diversion surgery are mostly suffering from tumorous diseases, so the collected cell sources may contain tumor cells. Other studies have shown that in vitro neuropathic bladder cells (smooth muscle and epithelial cells) have low contractile potential and low proliferation and differentiation ability [16]. Therefore, mesenchymal stem cells seem to be a good source of cells in bladder tissue engineering. However, since mesenchymal stem cells are not autologous bladder cells, they are easily altered by urinary toxicity, which may affect the success rate of tissue regeneration, but their advantage is that mesenchymal stem cells seem not to be affected by the initial pathology of the nervous system [17]. Studies have proved that the use of multiple cell co-transplantation can promote the regeneration of bladder tissue [18]. However, insufficient vascularization of regenerated bladder tissue is still a challenge for bladder tissue engineering. Zhao et al. [19] Isolated adipose-derived endothelial progenitor cells with high proliferation potential and angiogenic properties, pretreated with hypoxia to increase the activity of stem cells, and then combined porcine bladder acellular matrix with hypoxia-pretreated autologous adipose-derived endothelial progenitor cells were injected into the rat bladder reconstruction model at the same time, and its feasibility and the possibility of bladder blood vessel formation were evaluated, proving that hypoxia preconditioning can promote angiogenesis and tissue-engineered bladder function recovery.

Clinically, the ideal replacement bladder has complicated complications. The materials used for bladder reconstruction in the past included gelatin sponge, Japanese paper, formalin-preserved dog bladder, freeze-dried human dura mater, bovine pericardium, and small intestinal submucosa [20]. These biological materials provide a temporary scaffold for tissue growth, and over time they will reshape and degrade. A common complication in patients with bladder reconstruction is the gradual decrease in bladder capacity. Therefore, although these batteries can benefit some patients, they are rarely used in subsequent studies due to their poor long-term effects and the existence of complications. Be used again. To construct a tissue-engineered bladder, Bouhout et al. [21] Simulated an autologous bladder equivalent (VE). First, fibroblasts and epithelial cells were used for three-dimensional culture to obtain reconstructed VE; then they were cultured in a bioreactor, the reactor can provide a circulating pressure of up to 15 cm H2O, which is then rapidly reduced to achieve dynamically cultured VE (DCVE). The study found that the contour of the urothelium produced by dynamic culture was like that of the natural bladder. Permeability analysis showed that its contour was like that of the natural bladder and consistent with the basal cell tissue of the bladder. At the same time, it had the proper stretch ability during suture and processing. This new alternative method provides a new direction for regenerative medicine, which has similar properties to the bladder and can act as a urea barrier. These properties can significantly reduce inflammation, necrosis, and possible rejection. Studies have shown that, compared with cell-free scaffolds, cells are implanted into scaffolds (such as: bladder acellular matrix; polyacetic acid/polylactic acid-glycolic acid), the effect of cell growth is better [22-23]. In clinical trials, it was found that a cellular scaffold can only regenerate the urothelium but cannot regenerate smooth muscles to achieve the effect of restoring the contractile function of the bladder. Another study by Bouhout et al. [24] Found that the use of collagen-derived scaffolds in tissue-engineered three-dimensional spherical bladder models can mimic the natural shape of the bladder. Bladder mesenchymal cells are embedded in the scaffold, and epithelial cells are implanted on the surface. Therefore, the tension in the bladder mesenchymal cells and epithelial cells implanted in the model are like that in natural tissues. The three-dimensional spherical bladder model has the characteristics of highly mature urethral epithelium, collagen remodeling, and smooth muscle cells expressing myosin show a tendency to rearrange parallel to the surface of the cavity, which has characteristics comparable to natural tissues. This technology can be used for Partial replacement therapy for pathological bladder in the future.

The ideal tissue engineering, regeneration biomaterial should be able to make the mature epithelial cell layer uniformly and continuously attach to the surface of the bladder cavity and form multiple smooth muscle cell layers on the surface of the bladder cavity. It should also provide sufficient mechanical support to prevent renewal in the body. The tissue collapses prematurely before it forms. However, the implanted cells are suddenly exposed to an environment full of inflammatory mediators and activated immune responses, which will significantly affect their proliferation and function. Therefore, the constructed in vivo microenvironment should support the self-renewal, survival and differentiation of the implanted cell population and protect it from harmful factors [25]. Studies have found that the cocoon-like structure of the biomaterial matrix can reduce the damage to the implanted cell components, and the simple coupling of the immunomodulator and the biomaterial scaffold may turn it into a new "immunomodulator" biomaterial, thereby causing the host Active immune rejection [26]. In the future, if the paracrine signaling pathway and immune mechanism are understood more fully, the behavior of host immune cells can be guided more reasonably in clinical practice and appropriate regulation can be carried out, so that it can provide support for the uptake and integration of grafts into surrounding tissues.

## 6. Application of tissue engineering technology in urethral repair

The structure of the urethra is complicated, and it is susceptible to various injuries. The healing process often leads to the formation of scars. Autologous urine-derived stem cells and adipose-derived stem cells have great advantages in tissue engineering applications because of their noninvasiveness. They can be expanded in vitro and used in tissue engineering and three-dimensional bioprinting. However, if it is not sufficiently structured in the body, it may cause urine leakage or insufficient blood vessel formation. In addition, if it triggers an autoimmune reaction, it is easy to form a urethral stricture. By dynamically simulating the characteristics of contraction and expansion of the urethra in vitro, it helps the director play a key auxiliary role in the process of tissue differentiation in vitro. The reconstruction of a long urethra requires a long enough tubular graft, in which smooth muscle cells cover the inner and outer layers of the urothelium. Due to the activity of telomerase, if the material of mesenchymal stem cells is selected, their proliferation ability is strong and can be well Attaching to the surface of different scaffolds highlights certain advantages; when these mesenchymal stem cells differentiate into smooth muscle cells or epithelial cell lines, typical markers can be detected. Both urine-derived stem cells and bone marrow-derived stem cells can successfully differentiate into smooth muscle cell lines. However, the efficiency of producing epithelial cells in urine-derived stem cells is higher. The reason may be related to the source of urothelium [27-28]. The dynamic conditions of the director in the in vitro culture process are very critical. Because it simulates the spatial structure and environment, it has a positive impact on the proliferation of cells, the growth in the scaffold, and the maturation of engineered tissues [29-30]. The application of three-dimensional bioprinting technology makes it possible to prepare cell-loaded urethra with different polymer types and structural characteristics in vitro.

Zhang et al. [31] Used poly (ε-caprolactone) /poly (ε-caprolactone-collected), PCL-PLCL] polymer as a scaffold material to simulate the structure of rabbit urethra and mechanical properties, and use cell-loaded fibrin hydrogel to provide a microenvironment for cell growth. The results of the study show that the mechanical properties of the PCLPLCL (50:50) spiral stent are equivalent to that of the rabbit urethra. Evaluation of the biological activity of the cells in the bioprinted urethra showed that epithelial cells and smooth muscle cells remained more than 80% viable 7 days after printing. Both cell types can proliferate activity in the cell-carrying hydrogel and maintain specific biomarkers. These results provide a basis for further research on the three-dimensional bionics of the urethral tissue. The model also simulates the mechanical properties and cellular biological activity of the urethral tissue, as well as the possibility of using biofuel constructs in the biological model to implant the urethra in vivo.

Simoes et al. [32] Used biological scaffolds and derived soluble products to invent a method to produce decellularized urethral scaffolds by decellularization of porcine urethra. First staining with hematoxylineosin (HE), 4', 6-diamidino-2-phenylindole [2-(4-Amidinophenyl) -6-indolecarbamidine dihydrochloride, DAPI] staining and DNA quantification Method to evaluate the cell removal rate; then use immunofluorescence staining and calorimetric analysis kit to detect extracellular matrix protein; use human skeletal muscle myoblasts, muscle precursor cells and adipose-derived stromal vascular components to evaluate the regeneration of cellular urethral biofilms Vascularization; using mechanochemical decellularization method to remove about 93% of tissue DNA, basically retain the components and microstructure of the extracellular matrix, and finally achieve recellularization. It was confirmed by immunofluorescence and real-time fluorescence quantitative polymerase chain reaction that the skeletal muscle extracellular matrix promoted the formation of fibers and the expression of major skeletal muscle-related proteins and genes. This method can produce urethral biocomplexes that retain important extracellular matrix proteins and is easy to proliferate. This is the key first step in the preparation of tissue engineering technology based on urethral biocomplexes.

Clinically, urethral stricture is more common in men, so research mainly focuses on male urethral reconstruction. However, because the complications are difficult to control, the current commonly used surgical methods such as autogenous genital skin, bladder mucosa, oral mucosa replacement, etc. are not ideal. They often cause hair growth, stenosis, stone formation, and diverticulum. There is currently no standardized stent available for clinical use. However, in the clinical study of Raya-Rivera et al. [33], patients who need urethral reconstruction were evaluated, and the effectiveness of the tissue engineering urethra using the patient's own cells was considered. The research team selected 5 men with urethral defects as the research subjects and took tissue biopsy samples. The muscle cells and epithelial cells were expanded and inoculated on the polyglycolic acid-polylactide complex scaffold. The patients then used the tissue The engineered tubular urethra was reconstructed, and urinalysis, cystourethroscopy, cystourethrography, and blood flow measurement were performed at 3, 6, 12, 24, 36, 48, 60, and 72 months after surgery. At 3, 12 and 36 months, a series of endoscopic circular sampling biopsies were performed in different areas of the engineered urethra; the median follow-up time was 71 months (36-76 months). Anti-cytokeratin antibodies, actin, desmin and myosin antibodies confirmed the presence of epithelial cells and muscle cell lines in the culture. The median maximum urinary flow rate was 27.1 mL/s (normal range 16-28 mL/s). Continuous radiography and endoscopy showed that the urethra width was normal and there was no stricture. Urethral examination showed that the structure of the urethra was normal 3 months after transplantation. The results show that the tubular urethra can be used in a clinical environment and remain functional for up to 6 years, proving that the engineered urethra may be used in the future for patients who require complex urethral reconstruction. However, for adult long urethral strictures, such as those related to lichen sclerosis, it is still necessary to rely on autologous tissue replacement. At present, there is no other effective and clinically proven tissue engineering program for the clinical treatment of urethral strictures [34], Therefore, this field will also be a key breakthrough direction for future urethral tissue engineering research.

## 7. Conclusion

Tissue engineering technology is an emerging technology with broad prospects in the current medical field. Its potential is mainly manifested in a high degree of imitation of the cells, three-dimensional structure and environment required for the function of tissues and organs, which can maximize the restoration of their original functions after replacement., Even in the future can surpass the effect of its original function. Organisms, especially developing embryos, are typical self-assembly systems. Tissue and organogenesis are also completed through self-assembly. In the process of self-assembly, through the interaction of cell-cell and cell-extracellular matrix, the developing organism and its parts gradually differentiate and build into the final shape. Applying this characteristic to tissue engineering technology is an important idea for the development of tissue engineering. The success of designing and manufacturing functional structures and organs depends on the study of the principles of cell self-assembly and the ability of people to apply them to the clinic. There have been some successful cases [35], but there is no authoritative solution to meet the increasing demand for new regeneration technologies. Therefore, in the future, biological preparation methods (including but not limited to the field of tissue engineering) for reconstructing damaged tissues and organ functions will not only focus on the use of the body's regenerative capacity, but also the research of appropriate materials. Therefore, biomedicine and engineering are needed. Deeper integration. Nowadays, the main problems facing tissue engineering technology research are: the biocompatibility and mechanical properties of the scaffold, the proliferation of seed cells in the scaffold material does not meet the clinical requirements, and the immune inflammatory response causes scar formation and affects the transplantation effect. As an emerging technology, three-dimensional printing technology has high controllability of the printed structures and the choice of materials with good biocompatibility for printing, so its application in tissue engineering has great potential. With the emergence of new materials, the development and perfection of 3D printing technology, the continuous improvement and innovation of tissue engineering technology and concepts, and the continuous, in-depth understanding of cell interaction mechanisms, tissue engineering technology applied in clinical practice in the future will become a history of human medical treatment.

## References

1. Song L, Murphy S V, Yang B, et al. Bladder acellular matrix and its application in bladder augmentation. Tissue Eng Part B Rev, 2014, 20(2): 163-172.

2. Shafiee A, Atala A. Printing technologies for medical applications. Trends Mol Med, 2016, 22(3): 254-265.

3. Moon K H, Ko I K, Yoo J J, et al. kidney diseases and tissue engineering. Methods, 2016, 99: 112-119.

4. Setayeshmehr M, Esfandiari E, Rafieinia M, et al. Hybrid and composite scaffolds based on extracellular matrices for cartilage tissue engineering. Tissue Eng Part B Rev, 2019, 25(3): 202-224.

5. Song J J, Guyette J P, Gilpin S E, et al. Regeneration and experimental orthotopic transplantation of a bioengineered kidney. Nat Med, 2013, 19(5): 646-651.

6. Ko I K, Abolbashari M, Huling J, et al. Enhanced reendothelialization of acellular kidney scaffolds for whole organ engineering via antibody conjugation of vasculatures. Technology, 2014, 2(03): 243-253.

7. Lertkiatmongkol P, Liao D, Mei H, et al. Endothelial functions of platelet/endothelial cell adhesion molecule 1(CD31). Curr Opin Hematol, 2016, 23(3): 253-259.

8. Machiguchi T, Nakamura T. Nephron generation in kidney cortices through injection of pretreated mesenchymal stem celldifferentiated tubular epithelial cells. Biochem Biophys Res Commun, 2019, 518(1): 141-147.

9. Zhang J, Li K, Kong F, et al. Induced intermediate mesoderm combined with decellularized kidney scaffolds for functional engineering kidney. Tissue Engineering and Regenerative Medicine, 2019, 16(5): 501-512.

10. Orlando G, Booth C, Wang Z, et al. Discarded human kidneys as a source of ECM scaffold for kidney regeneration technologies. Biomaterials, 2013, 34(24): 5915-5925.

11. Koch H, Hammer N, Ossmann S, et al. Tissue engineering of ureteral grafts: preparation of biocompatible crosslinked ureteral scaffolds of porcine origin. Frontiers in Bioengineering and Biotechnology, 2015, 3: 89.

12. Zhao Z, Yu H, Xiao F, et al. Differentiation of adipose-derived stem cells promotes regeneration of smooth muscle for ureteral tissue engineering. J Surg Res, 2012, 178(1): 55-62.

13. Engel O, Petriconi R D, Volkmer B G, et al. The feasibility of ureteral tissue engineering using autologous veins: an orthotopic animal model with long term results. J Negat Results Biomed, 2014, 13(1): 1-9.

14. Lam Van Ba O, Aharony S., Loutochin O., et al Bladder tissue engineering: a literature review. Adv Drug Deliv Rev, 2015, 82-83: 31-37.

15. Shakhssalim N, Soleimani M, Dehghan M M, et al. Bladder smooth muscle cells on electrospun poly(ε-caprolactone)/poly(l-lactic acid) scaffold promote bladder regeneration in a canine model. Materials Science and Engineering: C, 2017, 75: 877-884.

16. Subramaniam R, Hinley J, Stahlschmidt J, et al. Tissue engineering potential of urothelial cells from diseased bladders. Journal of Urology, 2011, 186(5): 2014-2020.

17. Adamowicz J, Kloskowski T, Tworkiewicz J, et al. Urine is a highly cytotoxic agent: does it influence stem cell therapies in urology? Transplant Proc, 2012, 44(5): 1439-1441.

18. Sharma A K, Bury M I, Fuller N J, et al. Cotransplantation with specific populations of spina bifida bone marrow stem/progenitor cells enhances urinary bladder regeneration. Proc Natl Acad Sci USA, 2013, 110(10): 4003-4008.

19. Zhao Feng, Zhou Lliuhua, Xu Zhongle, et al. Hypoxiapreconditioned adipose-derived endothelial progenitor cells promote bladder augmentation. Tissue Eng Part A, 2019.

20. Adamowicz J, Pokrywczynska M, Vontelin V S, et al. Concise review: tissue engineering of urinary bladder; we still have a long way to go? Stem Cells Transl Med, 2017, 6(11): 2033-2043.

21. Bouhout S, Gauvin R, Gibot L, et al. Bladder substitute reconstructed in a physiological pressure environment. J Pediatr Urol, 2011, 7(3): 276-282.

22. Liao W, Yang S, Song C, et al. Tissue-engineered tubular graft for urinary diversion after radical cystectomy in rabbits. Journal of Surgical Research, 2013, 182(2): 185-191.

23. Basu J, Jayo M J, Ilagan R M, et al. Regeneration of native-like neourinary tissue from nonbladder cell sources. Tissue Eng Part A, 2012, 18(9-10): 1025.

24. Bouhout S., Chabaud S., Bolduc S. Collagen hollow structure for bladder tissue engineering. Materials Science and Engineering: C, 2019, 102: 228-237.

25. Nakayama K H, Luqia H, Huang N F. Role of extracellular matrix signaling cues in modulating cell fate commitment for cardiovascular tissue engineering. Adv Healthc Mater, 2014, 3(5): 628-641.

26. Vishwakarma A, Bhise N S, Evangelista M B, et al. Engineering immunomodulatory biomaterials to tune the inflammatory response. Trends Biotechnol, 2016, 34(6): 470-482.

27. Zhang Deying, Wei Guanghui, Li Peng, et al. Urine-derived stem cells: a novel and versatile progenitor source for cell-based therapy and regenerative medicine. Genes & diseases, 2014, 1(1): 8-17.

28. Qin D, Long Ting, Deng Junhong, et al. Urine-derived stem cells for potential use in bladder repair. Stem Cell Res Ther, 2014, 5(3): 69.

29. Wang Y, Fu Q, Zhao R Y, et al. Muscular tubes of urethra engineered from adipose-derived stem cells and polyglycolic acid mesh in a bioreactor. Biotechnol Lett, 2014, 36(9): 1909-1916.

30. Fu Q, Deng C L, Zhao R Y, et al. The effect of mechanical extension stimulation combined

with epithelial cell sorting on outcomes of implanted tissue-engineered muscular urethras. Biomaterials, 2014, 35(1): 105-112.

31. Zhang K., Fu Q., Yoo J., et al 3D bioprinting of urethra with PCL/PLCL blend and dual autologous cells in fibrin hydrogel: an in vitro evaluation of biomimetic mechanical property and cell growth environment. Acta Biomater, 2017, 50: 154-164.

32. Simoes I N, Vale P, Soker S, et al. Acellular urethra bioscaffold: decellularization of whole urethras for tissue engineering applications. Sci Rep, 2017, 7: 41934.

33. Raya-Rivera A, Esquiliano D R, Yoo J J, et al. Tissue-engineered autologous urethras for patients who need reconstruction: an observational study. Lancet, 2011, 377(9772): 1175-1182.

34. Chapple C. Tissue engineering of the urethra: where are we in 2019? World J Urol, 2019. https://doi.org/10.1007/s00345-019-02826-3.

35. Anthony A, Bauer S B, Shay S, et al. Tissue-engineered autologous bladders for patients needing cystoplasty. Lancet, 2006, 367(9518): 1241-1246.